# NICMOS Imaging of Candidate Extra-Solar Planetary Systems

Glenn Schneider (Steward Observatory, U. of Arizona) and the NICMOS IDT/EONS* Teams

* Near Infrared Camera and Multi-Object Spectrometer Instrument Definition / Environmments Of Nearby Stars Team: G. Schneider, U. AZ, E.E. Becklin, UCLA, B.A. Smith, U. HI, R.J. Terrile, JPL, R.I. Thompson, U. AZ, F.J. Low, U. AZ, A.J. Weinberger, UCLA, P.J. Lowrance, UCLA, M. Silverstone, U. AZ, C. Dumas, JPL, D.C. Hines, U. AZ, D.W. McCarthy, U. AZ, J.D. Kirkpatrick, IPAC, D.W. Koerner, U. PA

**ABSTRACT:** The connection between the formation and growth of planetesimals into larger bodies and the circumstellar debris disks around young and nascent stars from which they arise has been the subject of continual speculation, discussion and inquiry, since the original scenario for such evolution was first expanded upon by Laplace (1796) following the general proposition by Kant (1755). NICMOS coronagraphy has enabled both the direct imaging of such circumstellar disks seen in scattered light about more mature stars (other than β Pictoris) exhibiting morphologically resolved spatial features, and the detection of substellar companions (brown dwarfs and extra-solar giant planets) with masses extending into the Jovian domain for recently formed objects (< a few x $10^7$ years). Exploiting this capability, approximately 18% of the NICMOS IDTs observing time on *HST* was dedicated to addressing fundamental questions relating to the *ORIGINS* of extra-solar systems while exploring the Environments of Nearby Stars under the umbrella of our EONS project.

**Introduction:** Is there a continuity of companion objects across the sub-stellar mass-spectrum bridging the stellar main sequence into the planetary domain? In what sort of local environments will such objects form? At what distances will these objects be found from their primaries and how is this biased by the characteristics of the primary and companion objects and of the circumstellar regions? What implications will the discovery and characterization of such objects have for our understanding of formation mechanisms of extra-solar systems and their constituents? To begin answering these questions we conducted a survey of 85 stars selected as possible harbingers of circumstellar disks or sub-stellar companions in suite of coordinated observing programs:

• **A search for massive Jupiters (7226).** Observed 38 very young main-sequence stars with a mean distance of ~30 pc. The median age for candidates with well-established ages was ~90 Myrs with eight candidates as young as < 10 Myrs, including several members of the TW Hydrae association (TWA; d ~ 50pc). Sub-stellar companions to young stars are in higher luminosity phases and thus more readily detectable (L ~ $t^{-1.3}M^{2.24}$), and debris disks may still be present following the early epochs of planet-building.

• **A search for low mass companions to nearby M stars (7227).** Observed 27 M-dwarfs which are: a) very nearby (d < 6pc), with spectral types later than ~M3.5; b) young (age < 10 Myr) with (d < 25 pc); and c) spectrally the latest known (i.e., "ultra-cool" dwarfs later than ~M8.5), with some overlaps.

• **Dust disks around main sequence stars (7233).** Observed 18 primarily main-sequence stars with large IRAS excesses and other dust indicators at 1.1 and/or 1.6 μm. For some brighter stars we obtained multi-spectral images at 1.71μm (HCO2+C2 continuum), 2.04μm (methane band), and in the line regions of Paschen-α (1.87μm) and Bracket-γ (2.15μm).

• Separate programs obtained multicolor coronagraphic and polarimetric images of β Pictoris (7248), and high-resolution direct and coronagraphic imaging of HD 98800 (7232).

Here we present a sampling of results from our NICMOS coronagraphic imaging surveys, which we are now following up, where possible, with *HST*/SITS spectroscopy and with ground-based astrometry of the substellar companion candidates with the Keck Adaptive Optics system. With the upcoming installation of the NICMOS Cooling System in serving mission 3B and the advent of the resurrection of NICMOS in *HST* Cycle 10 we look forward to continuing our coronagraphic imaging surveys.

This work is based on observations with the NASA/ESA Hubble Space Telescope, obtained at STScI, which is operated by AURA, Inc., under NASA contract NAS2-6555 and supported by NASA grants NAG-03042 & GO-98.8176A to the NICMOS IDT and EONS teams.

**HR 4796A (TWA 11A) Dust Disk - Circumstellar Debris Ring[1].** On 15 March 1998 NICMOS obtained the first scattered light image of a circumstellar debris disk since the discovery of the Beta Pictoris disk by Smith and Terrile. The circumstellar dust in the HR 4796A disk is confined in a narrow ring 70 AU in radius and < 14 AU in width (FWHM). The dust ring is imaged azimuthally all around the central star in this NICMOS PSF-subtracted coronagraphic image. Scattered light from material near the minor axis of the ring ellipse, at an angular distance of 0.25" from the central star, is broadened by the 0.11" FWHM point spread function and "spills" out of the 0.3" radius coronagraph (white circle) allowing the ring to be seen in its entirety.

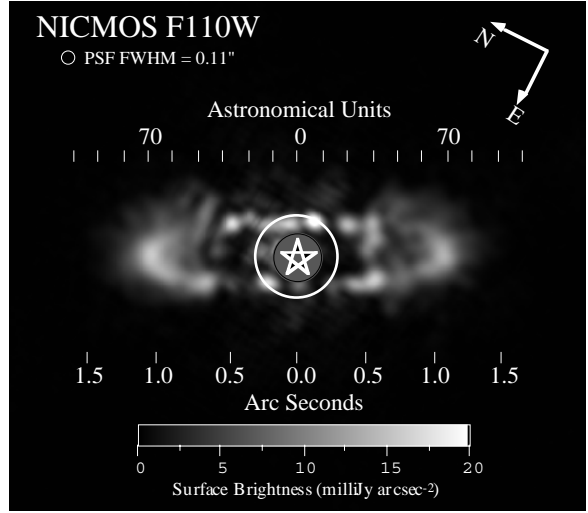

HR 4796A: d = 67 ± 3.5pc (Hipparcos), spec = A0V (Houck 1982) age = 8 ± 3Myr (Stauffer 1995)

```
GEOMETRY                    MORPHOLOGY              PHOTOMETRY
PA = 26.8° ± 0.6°           r = 70 AU               Flux @ 1.1µm = 12.8 ± 1.0mJy
i  = 73.1° ± 1.2°           Width < 14 AU           Flux @ 1.6µm = 12.5 ± 2.0mJy
a  = 1.05" ± 0.02"          "abrupt" truncation     H(F160W) = 12.35 (-0.19,+0.6)
                            "clear" @ r<50 AU       I(F110W) = 12.92 ± 0.08

OPTICAL DEPTH, τ_dust ~ L_disk/L*                   IMPLICATIONS
τ_dust(1.1µm) = 1.4±0.2x10⁻³                        • Mean particle size > few microns, debris
τ_dust(1.6µm) = 2.4±0.5x10⁻³                          origin and not trapped interstellar dust.
                                                    • Dynamical confinement of particles
Near-IR scattered flux in good agreement with absorption   may be due to the influence of one or
at visible wavelengths and rereadiation in the mid-IR.     more unseen bodies.
```

**HD 141569 - Circumstellar Disk and Gap[2].** The Herbig Ae/Be star HD 141569 (B9V, H=6.89, d = 99 ± 8pc, 10 Myr, 2.3 $M_{sun}$) was found to posess a circumstellar disk extending to a radius of at least 400 AU exhibiting a complex morphology including a 40 AU wide gap in its surface brightness profile at a radius of 250 AU. The disk, with a total flux density of 8±2 mJy beyond 0.6" (peak surface brightness 0.3mJy arcsec$^{-2}$ at 185 AU) is inclined to our line-of-site by 51°±3°. The intrinsic scattering function of the disk results in a brightness anisotropy in the ratio 1.5 ± 0.2 with the brighter side in the direction of forward scattering.

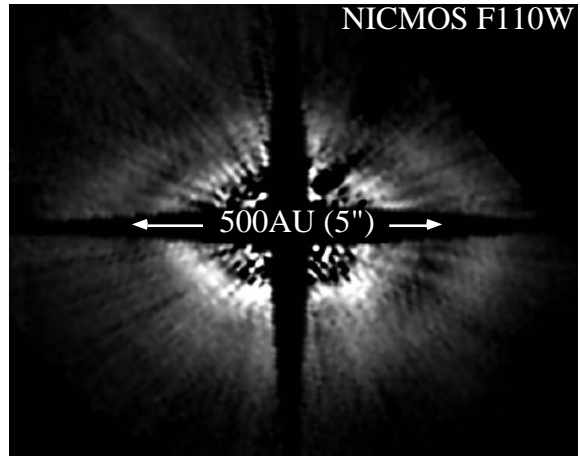

The region of the gap may be partialy cleared of material by an unseen co-orbital planetary companion. If so the width/radius ratio of the gap implies a planetary mass of ~ 1.3 Jupiters. This is consistant with our 2 Jupiter mass detection limit at this radius, where we also estimate the albedo, ω, to be 0.35 ± 0.05. HD 141569"A" may be part of a triple system with two likely "double" nearby stars (d = 7.6" and 8.9", 1.3" separation). If so they undoubtedly influence the dynamics of the disk.

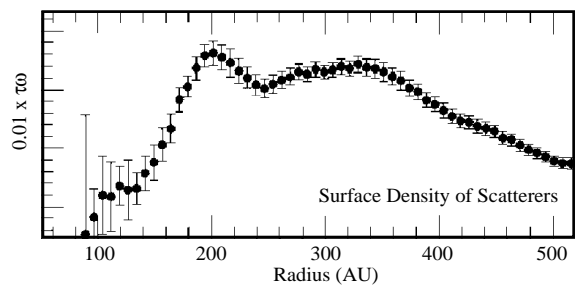

**ROSAT 116B (TWA 6B) - Giant Planet Candiate[3].** Direct imaging of young Jovian-mass planets is enabled by NICMOS coronagraphy as demonstrated in this S/N=50 1.6 µm (F160W) image of a candidate extra-solar giant planet companion to ROSAT 116 (H=6.9, K7V). This X-ray source is a member of the TW Hydra association, a loose stellar association of, perhaps, two dozen now-identified members, all ~ 10 Myr of age and comprising the nearest site of recent star formation to the Earth (d ≈ 50 pc). The H=20.1 putative companion is 2.5" (125 AU projected distance) from its 13.2 magnitude brighter primary. In a follow-up NICMOS camera 1 observation at 0.9µm (F090M) the object was not seen to a limiting magnitude of > 22, indicative of a very red source (if not a highly reddened background object). If the object is physically associated and coeval with ROSAT 116, its absolute H magnitude suggests an effective surface temperature of ~ 800K (for a surface gravity of $7.5 \times 10^4$) and a mass of ~ 1 Jupiter based upon evolutionary models by Burrows (1997). We will obtain a STIS 0.75 - 1.0 µm spectrum of the object during HST Cycle 8 to ascertain its physical nature. We are attempting to use the Keck AO system to establish (or reject) its physical association with ROSAT 116. Very preliminary proper motion measures indicate TWA6B is 3.9σ from the expected position of a background star, but 1.6σ from the expected position for a planet. Further work is needed to improve the relative astrometry.

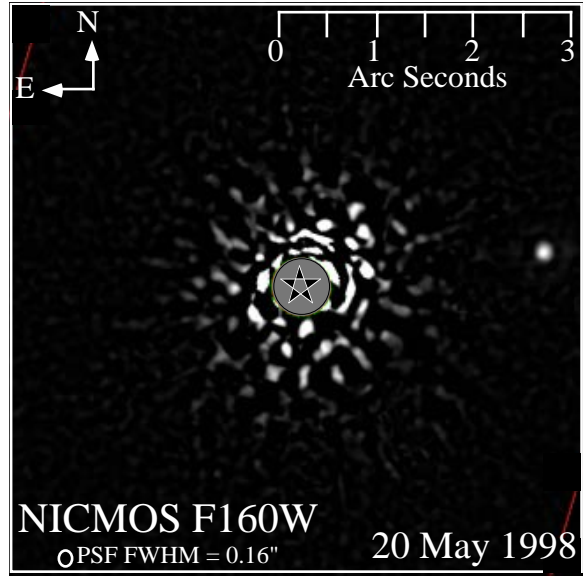

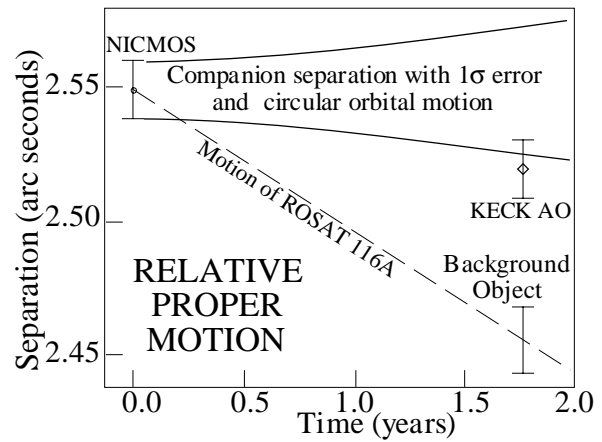

**TW Hydrae (TWA 1) Face On Circumstellar Disk[4]**. TW Hydrae (V=11.17, H=7.37, K7V), a classical T-Tauri star, is the archetypal member of the young stellar association which bears its name. TW Hya was found to harbor an optically thick face-on disk (r ≈ 190 AU) seen in NICMOS F110W and F160W coronagraphic images (and also by Stapelfeldt with WFPC-2), which is fit very well at both wavelengths with an $r^{-2.6}$ power law. Areal scattering profiles, in both colors, corrected for the color of the star reveal a break in the surface density of scatterers at R ≈ 100 AU which may be indicative of sculpting of the disk grains.

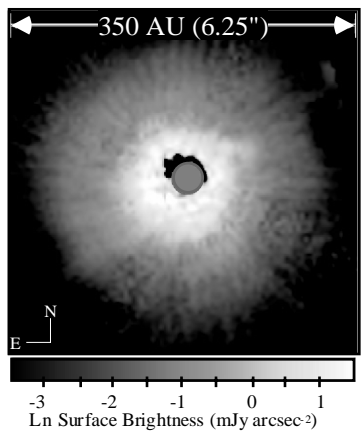

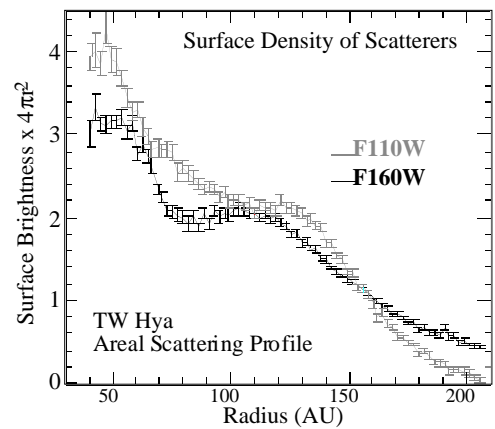

**HD 98800B (TWA 4B) - Planetary Debris System[5].** Quantitative estimates, and limits, on the geometry, size (r = 4.5 AU), albedo (< 0.3) and total mass (~ 0.6 $M_{earth}$) of the Planetary Debris System (PDS) around HD 98800B are ascertained from the non-detection of its highly emissive thermal IR disk in scattered light. A plausible model of the system is suggested from spectral energy distributions in combination with the luminosities of the individual stellar components (as determined by NICMOS), and that of the PDS as previously determined at long wavelengths. Remarkable similarities are found between the PDS and the debris system around the Sun as it could have appeared a few million years after its formation.

**Brown Dwarf Companions.** Recently, many old isolated "field" brown dwarfs have been discovered from large ground-based surveys such as the 2-Micron All-Sky and Sloan Digital Sky surveys and subsequently confirmed spectroscopically. Indeed, it has been suggested that brown dwarfs, once considered somewhat exotic objects, may be as common as stars. Yet, the companion mass fraction of these substellar objects which occupy a niche in the mass function between planets and stars remains largely unknown. How common are they? Do companion brown dwarfs form in a process more like planets than stars? How does the presence of a brown dwarf companion effect the evolution of a newly-forming solar system? Our small NICMOS survey begins to address these fundamental questions but many more objects, such as those we present below, will have to be discovered and studied before they are fully answered.

| PRIMARY | HR 7329A[6] | TWA 5A[7] |
|---|---|---|
| Distance | 47 pc | ~ 50 pc |
| Age | ~ 40 Myr | ~ 10 Myr |
| Magnitude | H=5.05 | H=7.2, I=8.8, K=6.8 |
| MKL Class | A0V | M1.5V |
| COMPANION | HR 7329B | TWA 5B |
| H mag | 11.90 | 2.14 |
| H abs | 8.54 | 8.41 |
| Colors | ----- | I-J=3.2, I-H=3.7, I-K=4.4 |
| MKL Class | M7.5V | M9V |
| Phot. Spec. | ----- | M8.5V |
| BC(H) | 2.69 | 2.80 |
| Teff | 2700K | 2600K |
| Luminosity | 0.0026 $L_{sun}$ | 0.0021 $L_{sun}$ |
| Mass | ~ 40 $M_{jup}$ | ~ 2 $M_{jup}$ |
| Separation | 4.17" | 1.96" |
| Proj. Dist. | 196 AU | 98 AU |
| PA (E from N) | 166.8° | 358.9° |
| Companionship[†] | 99.998% | 99.94% |

[†]Probabilities from space density surveys and colors. Confirmations via proper motions to be established.

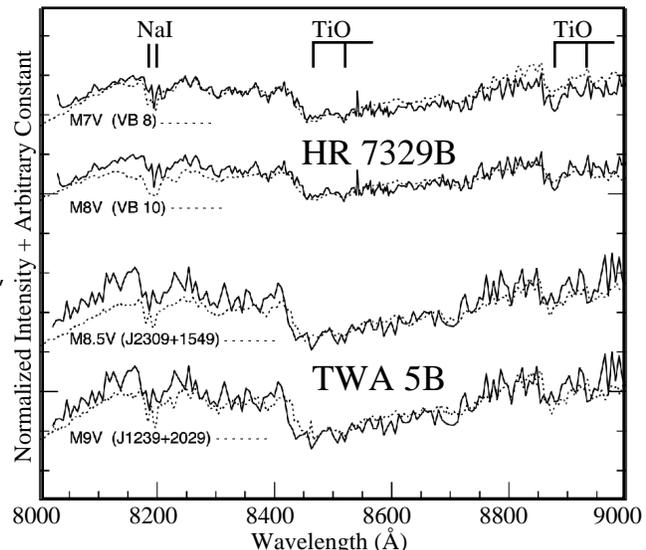

Young, and therefore hot, brown dwarfs have spectra resembling M dwarf stars of the same photospheric temperatures. Here we compare our recently acquired STIS spectra of TWA 5B and HR 7329B with several late M stars.

**REFERENCES:**

[1]Schneider, G., Smith, B.A., Becklin, E.E., Koerner, D.W., Meier, R., Hines, D.C., Lowrance, P.J., Terrile, R.J., Thompson, R.I., & Rieke, M., 1999, ApJ, 513, L127

[2]Weinberger, A.J., Becklin, E.E., Schneider, G., Smith, B.A., Lowrance, P.J., Silverstone, M.D., and Zuckerman, B., 1999, ApJ, 522, L53.

[3]Schneider, G., Becklin E.E., Lowrance, P., and Smith, B.A, 2000, APS Conference Series, in press.

[4]Weinberger, A.J., Becklin E.E., and Schneider, G., 2000, APS Conference Series, in press.

[5]Low, F.J., Hines, D.C., and Schneider, G., 1999, , ApJ, 520, L45.

[6]Lowrance, P.J., McCarthy, C., Becklin, E.E., Zuckerman, B.M., Schneider, G., Webb., R.A., Hines, D.C., Kirkpatrick, J.D., Koerner, D.W., Low, F.J., Meier, R., Rieke, M., Smith, B.A., Terrile, R.J., & Thompson, R.I., 1999, ApJ, 512 , L69.

[7]Lowrance, P.J., Becklin, E.E., Schneider, G., Kirkpatrick, J.D., Zuckerman, B., Plait, P., Malumuth, E., Heap, S.R., Weinberger, A.J., Smith, B.A., Terrile, R.J., Schultz, A.B., and Hines, D.C., ApJ, in press.

Original poster display available at: http://nicmosis.as.arizona.edu:8000/POSTERS/HST_DECADE10_POSTER.jpg